\documentclass[a4paper]{jpconf}
\usepackage{graphicx}
\usepackage{url}
\begin{document}
\title{Particle Physics with the Cosmic Microwave Background with SPT-3G}


\author{
J.~S.~Avva\textsuperscript{1}, \normalfont P.~A.~R.~Ade\textsuperscript{2},
Z.~Ahmed\textsuperscript{3,4},
A.~J.~Anderson\textsuperscript{5,6},
K.~Aylor\textsuperscript{7}, 
R.~Basu Thakur\textsuperscript{6} , 
A.~N.~Bender\textsuperscript{8,6}, 
B.~A.~Benson\textsuperscript{5,6,9}, 
L.~E.~Bleem\textsuperscript{8,6}, 
S.~Bocquet\textsuperscript{10,8}, 
L.~Bryant\textsuperscript{11} , 
J.~E.~Carlstrom\textsuperscript{6,11,12,8,9} , 
F.~W.~Carter\textsuperscript{8,6} , 
T.~W.~Cecil\textsuperscript{8} , 
C.~L.~Chang\textsuperscript{8,6,9} , 
T.~M.~Crawford\textsuperscript{6,9} , 
A.~Cukierman\textsuperscript{1} , 
T.~de~Haan\textsuperscript{1} , 
J.~Ding\textsuperscript{14} , 
M.~A.~Dobbs\textsuperscript{13,15} , 
S.~Dodelson\textsuperscript{16} , 
D.~Dutcher\textsuperscript{6,12} , 
W.~Everett\textsuperscript{17} , 
K.~R.~Ferguson\textsuperscript{18} , 
A.~Foster\textsuperscript{19} , 
J.~Gallicchio\textsuperscript{6,20} , 
A.~E.~Gambrel\textsuperscript{6} , 
R.~W.~Gardner\textsuperscript{11} , 
J.~C.~Groh\textsuperscript{1} , 
S.~Guns\textsuperscript{1} , 
N.~W.~Halverson\textsuperscript{17,21} , 
A.~H.~Harke-Hosemann\textsuperscript{8,22} , 
N.~L.~Harrington\textsuperscript{1} , 
J.~W.~Henning\textsuperscript{8,6} , 
G.~ P.~Holder\textsuperscript{15,23} , 
W.~L.~Holzapfel\textsuperscript{1} , 
D.~Howe\textsuperscript{24} , 
N.~Huang\textsuperscript{1} , 
K.~D.~Irwin\textsuperscript{3,25,4} , 
O.~B.~Jeong\textsuperscript{1} , 
M.~Jonas\textsuperscript{5} , 
A.~Jones\textsuperscript{24} , 
T.~S.~Khaire\textsuperscript{14} , 
L.~Knox\textsuperscript{7} , 
A.~M.~Kofman\textsuperscript{22} , 
M.~Korman\textsuperscript{19} , 
D.~L.~Kubik\textsuperscript{5} , 
S.~Kuhlmann\textsuperscript{8} , 
C.-L.~Kuo\textsuperscript{3,25,4} , 
A.~T.~Lee\textsuperscript{1,26} , 
A.~E.~Lowitz\textsuperscript{6} , 
S.~S.~Meyer\textsuperscript{6,11,12,9} , 
D.~Michalik\textsuperscript{24} , 
J.~Montgomery\textsuperscript{13} , 
A.~Nadolski\textsuperscript{22} , 
T.~Natoli\textsuperscript{6,9} , 
H.~Nguyen\textsuperscript{5} , 
G.~I.~Noble\textsuperscript{13} , 
V.~Novosad\textsuperscript{14} , 
S.~Padin\textsuperscript{6} , 
Z.~Pan\textsuperscript{6,12} , 
P.~Paschos\textsuperscript{11} , 
J.~Pearson\textsuperscript{14} , 
C.~M.~Posada\textsuperscript{14} , 
W.~Quan\textsuperscript{6,12} , 
S.~Raghunathan\textsuperscript{18} , 
A.~Rahlin\textsuperscript{5,6} , 
C.~L.~Reichardt\textsuperscript{27} , 
D.~Riebel\textsuperscript{24} , 
J.~E.~Ruhl\textsuperscript{19} , 
J.T.~Sayre\textsuperscript{17} , 
E.~Shirokoff\textsuperscript{6,9} , 
G.~Smecher\textsuperscript{28} , 
J.~A.~Sobrin\textsuperscript{6,12} , 
A.~A.~Stark\textsuperscript{29} , 
J.~Stephen\textsuperscript{11} , 
K.~T.~Story\textsuperscript{3,25} , 
A.~Suzuki\textsuperscript{26} , 
K.~L.~Thompson\textsuperscript{3,25,4} , 
C.~Tucker\textsuperscript{2} , 
K.~Vanderlinde\textsuperscript{30,31} , 
J.~D.~Vieira\textsuperscript{22,23} , 
G.~Wang\textsuperscript{8} , 
N.~Whitehorn\textsuperscript{18} , 
W.~L.~K.~Wu\textsuperscript{6} , 
V.~Yefremenko\textsuperscript{8} , 
K.~W.~Yoon\textsuperscript{3,25,4} , 
M.~R.~Young\textsuperscript{31} 
}
\medskip
\textsuperscript{1}{Department of Physics, University of California, Berkeley, CA, 94720, USA} \\
\textsuperscript{2}{School of Physics and Astronomy, Cardiff University, Cardiff CF24 3YB, United Kingdom} \\
\textsuperscript{3}{Kavli Institute for Particle Astrophysics and Cosmology, Stanford University, 452 Lomita Mall, Stanford, CA, 94305, USA} \\
\textsuperscript{4}{SLAC National Accelerator Laboratory, 2575 Sand Hill Road, Menlo Park, CA, 94025, USA} \\
\textsuperscript{5}{Fermi National Accelerator Laboratory, MS209, P.O. Box 500, Batavia, IL, 60510, USA} \\
\textsuperscript{6}{Kavli Institute for Cosmological Physics, University of Chicago, 5640 South Ellis Avenue, Chicago, IL, 60637, USA} \\
\textsuperscript{7}{Deptartment of Physics, University of California, One Shields Avenue, Davis, CA 95616, USA} \\
\textsuperscript{8}{High-Energy Physics Division, Argonne National Laboratory, 9700 South Cass Avenue., Argonne, IL, 60439, USA} \\
\textsuperscript{9}{Department of Astronomy and Astrophysics, University of Chicago, 5640 South Ellis Avenue, Chicago, IL, 60637, USA} \\
\textsuperscript{10}{Ludwig-Maximilians-Universit{\"a}t, Scheiner- str. 1, 81679 Munich, Germany} \\
\textsuperscript{11}{Enrico Fermi Institute, University of Chicago, 5640 South Ellis Avenue, Chicago, IL, 60637, USA} \\
\textsuperscript{12}{Department of Physics, University of Chicago, 5640 South Ellis Avenue, Chicago, IL, 60637, USA} \\
\textsuperscript{13}{Department of Physics and McGill Space Institute, McGill University, 3600 Rue University, Montreal, Quebec H3A 2T8, Canada} \\
\textsuperscript{14}{Materials Sciences Division, Argonne National Laboratory, 9700 South Cass Avenue, Argonne, IL, 60439, USA} \\
\textsuperscript{15}{Canadian Institute for Advanced Research, CIFAR Program in Gravity and the Extreme Universe, Toronto, ON, M5G 1Z8, Canada} \\
\textsuperscript{16}{Department of Physics, Carnegie Mellon University, Pittsburgh, Pennsylvania, 15312, USA} \\
\textsuperscript{17}{CASA, Department of Astrophysical and Planetary Sciences, University of Colorado, Boulder, CO, 80309, USA } \\
\textsuperscript{18}{Department of Physics and Astronomy, University of California, Los Angeles, CA, 90095, USA} \\
\textsuperscript{19}{Department of Physics, Center for Education and Research in Cosmology and Astrophysics, Case Western Reserve University, Cleveland, OH, 44106, USA} \\
\textsuperscript{20}{Harvey Mudd College, 301 Platt Boulevard., Claremont, CA, 91711, USA} \\
\textsuperscript{21}{Department of Physics, University of Colorado, Boulder, CO, 80309, USA} \\
\textsuperscript{22}{Department of Astronomy, University of Illinois at Urbana-Champaign, 1002 West Green Street, Urbana, IL, 61801, USA} \\
\textsuperscript{23}{Department of Physics, University of Illinois Urbana-Champaign, 1110 West Green Street, Urbana, IL, 61801, USA} \\
\textsuperscript{24}{University of Chicago, 5640 South Ellis Avenue, Chicago, IL, 60637, USA} \\
\textsuperscript{25}{Deptartment of Physics, Stanford University, 382 Via Pueblo Mall, Stanford, CA, 94305, USA} \\
\textsuperscript{26}{Physics Division, Lawrence Berkeley National Laboratory, Berkeley, CA, 94720, USA} \\
\textsuperscript{27}{School of Physics, University of Melbourne, Parkville, VIC 3010, Australia} \\
\textsuperscript{28}{Three-Speed Logic, Inc., Vancouver, B.C., V6A 2J8, Canada} \\
\textsuperscript{29}{Harvard-Smithsonian Center for Astrophysics, 60 Garden Street, Cambridge, MA, 02138, USA} \\
\textsuperscript{30}{Dunlap Institute for Astronomy \& Astrophysics, University of Toronto, 50 St. George Street, Toronto, ON, M5S 3H4, Canada} \\
\textsuperscript{31}{Department of Astronomy \& Astrophysics, University of Toronto, 50 St. George Street, Toronto, ON, M5S 3H4, Canada} \\

\ead{javva@berkeley.edu}

\begin{abstract}
The cosmic microwave background (CMB) encodes information about the content and evolution of the universe. The presence of light, weakly interacting particles impacts the expansion history of the early universe, which alters the temperature and polarization anisotropies of the CMB. In this way, current measurements of the CMB place interesting constraints on the neutrino energy density and mass, as well as on the abundance of other possible light relativistic particle species. We present the status of an on-going 1500 sq. deg. survey with the SPT-3G receiver, a new mm-wavelength camera on the 10-m diameter South Pole Telescope (SPT). The SPT-3G camera consists of 16,000 superconducting transition edge sensors, a 10x increase over the previous generation camera, which allows it to map the CMB with an unprecedented combination of sensitivity and angular resolution. We highlight projected constraints on the abundance of sterile neutrinos and the sum of the neutrino masses for the SPT-3G survey, which could help determine the neutrino mass hierarchy.
\end{abstract}

\section{Introduction}
The South Pole Telescope (SPT) is a 10-meter telescope located at the Amundsen-Scott South Pole Station designed to make high-resolution millimeter-wavelength measurements of the cosmic microwave background (CMB). The SPT-3G camera on the SPT was deployed in the 2016-2017 austral season. It consists of over 16,000 transition-edge sensors, a 10x increase in detectors over the previous-generation camera on the SPT \cite{benson2014spt} \cite{bender2018year}. These detectors are polarization-sensitive and span three frequencies -- 90, 150, and 220 GHz. These  significant improvements will allow the SPT-3G camera to map the polarization and temperature anisotropies of the CMB with unprecedented depth ($\sim$ 2 $\mu$K in temperature at 150 GHz), resolution (1 arcmin), and sky coverage (1500 deg$^2$). 

The CMB provides a unique way to probe particle physics since it was emitted in the early universe.  First, the anisotropies provide a snapshot at $\sim$380,000 years after the Big Bang. In this period, neutrinos and other light relics comprised around 10$\%$ of the energy density of the universe, as opposed to $<$ 1$\%$ now. Small-scale anisotropies are particularly affected by this radiation because small-scale structure is exponentially damped by an increased energy-to-mass ratio in the early universe. This allows for a sensitive measurement of $N_{\textup{eff}}$, the parameter that refers to the effective number of neutrino species \cite{weinberg2008cosmology}. The measurement of $N_{\textup{eff}}$ is described in Section 2. Second, CMB photons have been traveling through the universe for billions of years. The intervening matter between the surface of last scattering and the telescopes that observe the CMB affects the CMB photons, for example by scattering and gravitational lensing. The change to large-scale structure over time is thus imprinted in the CMB measurements that are made today. In this way, the large-scale structure, and therefore the effect of the sum of the neutrino masses on the expansion history of the universe, can be studied \cite{bond1983collisionless} \cite{lesgourgues2005cmb}. The measurement of $\sum m_\nu$ is described in Section 3.


\section{Light Relics and Neutrinos in the Early Universe}

A particle species in chemical equilibrium with the Standard Model particles at any point in the early universe contributes to the universe's total energy density. The fractional amounts of the energy density of the early universe that are composed of matter, dark energy, or radiation all change the expansion rate of the universe. The net effect of this is observed in the CMB temperature and polarization power spectra. The power spectra of the CMB have a series of peaks and troughs that decrease in amplitude at small angular scales and are the products of acoustic oscillations and photon diffusion damping \cite{kamionkowski1997statistics} \cite{zaldarriaga1997all}. The locations of the peaks and the shape of the damping tail determine the parameter $N_{\textup{eff}}$, the effective number of neutrino species \cite{hou2013massless}. The Standard Model predicts a value of  $N_{\textup{eff}}$  = 3.046. The difference between the Standard Model prediction and the naive prediction of 3 stems from entropy transferred to neutrinos during electron-positron annihilation due to neutrinos not being fully decoupled from the early universe plasma. Additional light relativistic particles contribute a change to $N_{\textup{eff}}$ related to the point at which they decoupled from the primordial plasma (related to their masses and cross-sections). This potential extra energy contribution to $N_{\textup{eff}}$ could be due to a sterile neutrino, axion-like particles, a matter/antimatter asymmetry in the neutrino sector, or any other possible light relic in the early universe \cite{abazajian2016cmb}. In Figure 1, contributions from additional light relics are shown as functions of their spin and freeze-out temperature. As seen in the figure, for a $\Lambda$CDM + $Y_{\textup{p}}$ (the primordial helium abundance) + $N_{\textup{eff}}$ cosmology, Planck has measured the current best limit, a change in $N_{\textup{eff}}$ of $\Delta N_{\textup{eff}}$ = 0.19 at 1$\sigma$ \cite{pl19}, which rules out particles that decoupled after quarks became confined to hadrons (the QCD phase transition). This corresponds to many more degrees of freedom in the early universe plasma, and consequentially smaller changes in $N_{\textup{eff}}$ for additional particle species. With SPT-3G's sensitivity of $\Delta N_{\textup{eff}}$ = 0.1 at 1$\sigma$, it is projected to be one of the first experiments which will measure beyond the QCD phase transition, placing interesting constraints on new physics.

\section{Measuring Neutrino Mass from Clusters and Lensing}

\begin{figure}[h]
\begin{minipage}{17pc}
\includegraphics[trim=140 30 150 30,clip,width=17pc]{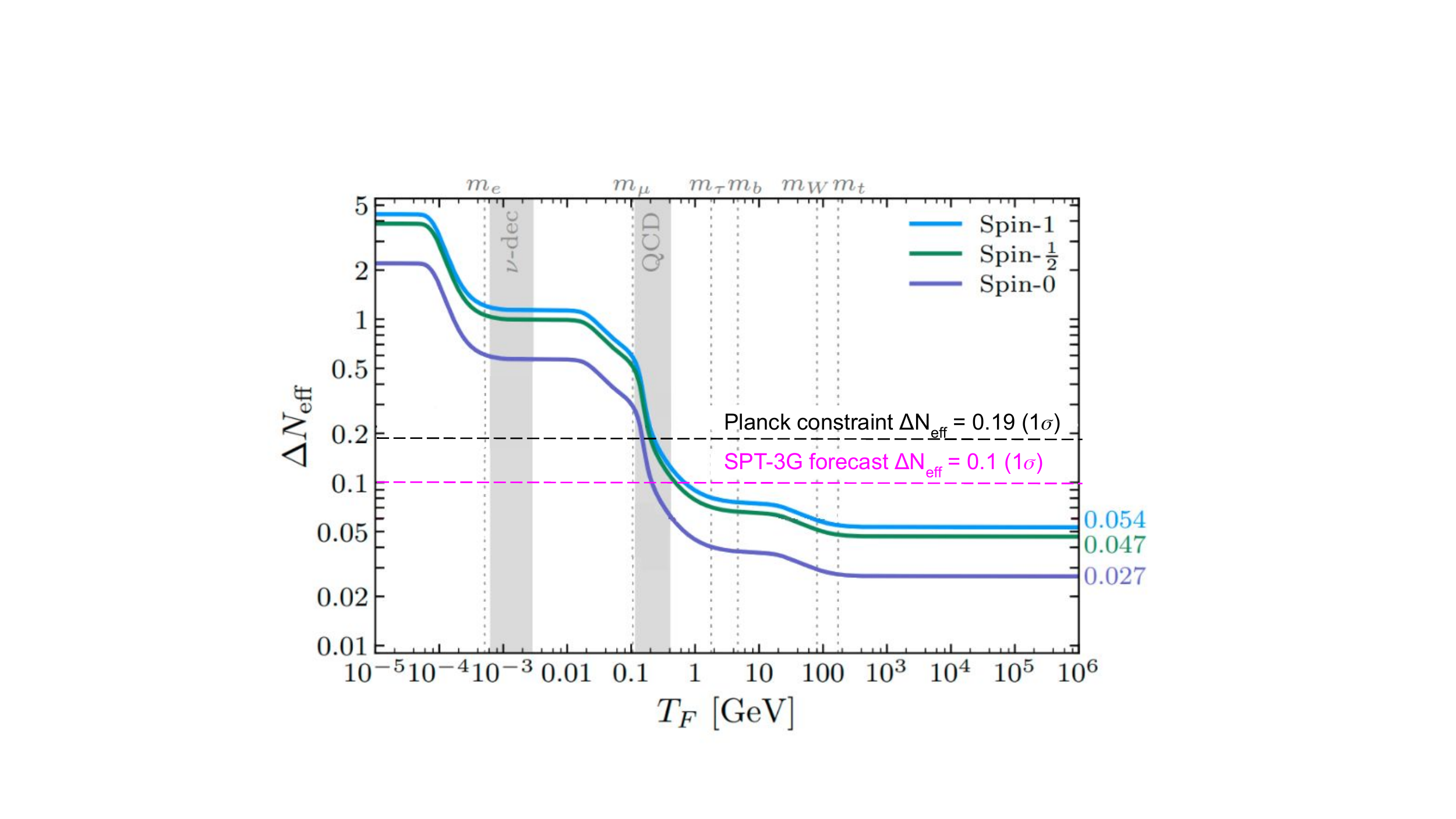}
\caption{\label{label}Contributions to $N_{\textup{eff}}$ from a single, massless, particle that froze-out from the Standard Model particles at temperature $T_F$. Figure adapted from Figure 10 of \cite{abazajian2019cmb}. The black dashed line shows the current best constraint on $\Delta N_{\textup{eff}}$ from $Planck$ \cite{pl19}, and the magenta line is the forecast for SPT-3G, showing that SPT-3G could go beyond the QCD phase transition.}
\end{minipage}\hspace{2pc}%
\begin{minipage}{15pc}
\includegraphics[width=15pc]{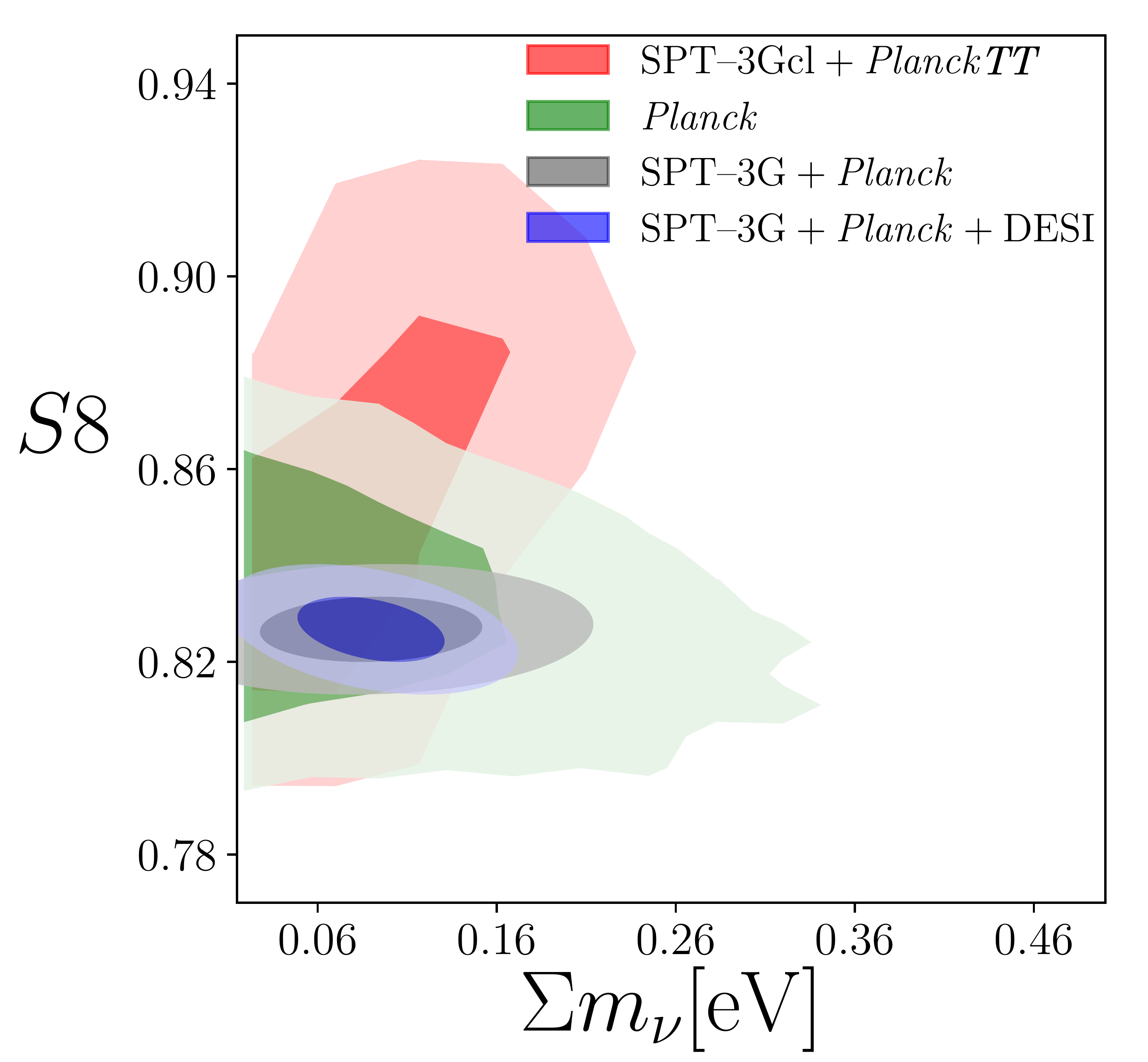}
\caption{\label{label}Projected 68 and 95$\%$ confidence intervals for $S8$ vs. $\sum m_\nu$ determined from: $Planck$ TT+TE+EE+lowE CMB
\cite{pl19} (green), the forecasted
improvement adding first SPT-3G CMB (gray) and then
DESI BAO data \cite{collaboration2018desi} (purple), and the forecasted constraints
from $Planck$ TT+lowE with the SPT-3G cluster data (red).}
\end{minipage} 
\end{figure}

In addition to being a snapshot of the early universe, the CMB also provides a probe of the evolution of cosmic structure over time. As neutrinos free-stream, they suppress small-scale structure as described in the previous section. This suppression is caused by the particles being ultra-relativistic. As the universe ages and they cool, they slow, changing their effect on structure. Combining the measurement of early structure from the preceding section with any of a number of measurements of more recent times thus provides an opportunity to measure the neutrinos' velocity as a function of energy, and thus their mass \cite{bond1983collisionless} \cite{lesgourgues2005cmb}. This is a complementary approach to terrestrial neutrino experiments -- the CMB measures $\sum m_\nu$, as opposed to the differences between the neutrino flavors. Due to SPT-3G's high angular resolution, the data will actually be able to provide two separate constraints of $\sum m_\nu$: one from galaxy clusters and one from CMB lensing. These two methods have different systematic uncertainties, so this provides a compelling cross-check for $\sum m_\nu$ constraints from cosmological data.

First, the abundance of galaxy clusters over time traces large-scale structure. As CMB photons pass through the hot gaseous regions of galaxy clusters, they can scatter off high-energy electrons. This imprints a characteristic spectral distortion called the Sunyaev-Zeldovich effect \cite{sunyaev1970spectrum} in CMB maps at the location of galaxy clusters. In this way, large-scale structure over time can be characterized by the abundance of galaxy clusters as a function of mass and redshift. 

Second, lensing also probes large-scale structure.  The intervening matter between us and the surface of last scattering distorts the CMB spectrum through lensing. The amplitude of the lensing signal is proportional to the amount of large-scale structure in the universe and as a result can be used to characterize the effect of massive neutrinos.  

These two measurements constrain $\sum m_\nu$ on a similar order of magnitude as displayed in Figure 2. This figure shows the joint confidence intervals for the sum of the neutrino masses and a parameter describing the combination of the amplitude of the matter power spectrum and the matter energy density. The quantity $S8$ is defined by the amplitude of the matter power spectrum ($\sigma_8$) and the sum of the cold dark matter, baryon, and massive neutrino energy densities ($\Omega_m$) via the relationship $S8 = \sigma_8 \sqrt{\Omega_m/0.3}$. The constraint from SPT-3G galaxy clusters is forecast to be $\sigma(\sum m_\nu )\simeq$ 0.055 eV and $\sigma(\sum m_\nu )\simeq$ 0.06 eV from CMB lensing (SPT-3G + $Planck$). Adding in the Dark Energy Spectroscopic Instrument (DESI) baryon acoustic oscillations (BAO) data \cite{collaboration2018desi}, the SPT-3G forecast improves to $\sigma(\sum m_\nu )\simeq$ 0.038 eV. Currently, neutrino oscillation experiments limit $\sum m_\nu \geq$ 0.067 eV. The inverted hierarchy dictates that there are two neutrinos with $m_\nu \geq$ 0.049 eV, which would make $\sum m_\nu \geq $ 0.107 eV \cite{particle2014review}.  Assuming the normal hierarchy and the minimum value of $\sum m_\nu =$ 0.067 eV, SPT-3G could provide evidence against the inverted hierarchy. Conversely, assuming the inverted hierarchy, SPT-3G could rule out $\sum m_\nu =$ 0 at 2$\sigma$.

\section{Summary}
 Using the polarization and temperature power spectra, SPT-3G will provide a 1$\sigma$ constraint of $\Delta N_{\textup{eff}}$ = 0.1, probing new parameter space for light relic particles in the early universe. Between the two systematically independent metrics of galaxy clusters and CMB lensing, SPT-3G will set a constraint of $\sigma(\sum m_\nu )\simeq$ 0.038 eV, a 30$\%$ improvement over $Planck$. Studying the CMB using SPT-3G data provides a cutting-edge and complementary window into the Standard Model and beyond.

\ack The South Pole Telescope is supported by the National Science Foundation (NSF)
through grant PLR-1248097.  The author acknowledges support from the National Science Foundation Graduate Research Fellowship under Grant No. DGE 1752814. Partial support is also provided by the NSF Physics Frontier Center grant PHY1125897 to the Kavli Institute of Cosmological Physics at the University of Chicago, and the Kavli Foundation and the Gordon and Betty Moore Foundation grant GBMF 947. Work at Argonne National Laboratory,
including Laboratory Directed Research and Development support and use of the Center for Nanoscale Materials, a U.S. Department of Energy, Office of Science (DOE-OS) user facility, was supported under Contract No. DE-AC02-06CH11357. We acknowledge R. Divan, L. Stan, C.S. Miller, and V. Kutepova for supporting our work in the Argonne Center for Nanoscale Materials. Work at Fermi National Accelerator Laboratory, a DOE-OS, HEP User Facility managed by the Fermi Research Alliance, LLC, was supported under Contract No. DE-AC02-07CH11359. NWH acknowledges support from NSF CAREER grant AST-0956135.
The McGill authors acknowledge funding from the Natural Sciences and Engineering Research Council of
Canada, Canadian Institute for Advanced Research, and the Fonds de recherche du Qubec Nature et technologies. JV acknowledges support from the Sloan Foundation.

\section*{References}
\bibliography{iopart-num}

\end{document}